\documentclass[preprint2]{aastex}

\shortauthors{Wanajo et al.}
\shorttitle{$r$-PROCESS IN SUPERNOVAE}

\begin{document}

\title{THE $r$-PROCESS IN SUPERNOVAE: IMPACT OF NEW MICROSCOPIC
MASS FORMULAS}

\author{\sc Shinya Wanajo\footnote{Department of Physics, Sophia University,
       7-1 Kioi-cho, Chiyoda-ku, Tokyo, 102-8554, Japan;
       wanajo@sophia.ac.jp, n\_itoh@sophia.ac.jp},
Stephane Goriely\footnote{Institut d'Astronomie et d'Astrophysique, C.P.
226, Universit\'e Libre de Bruxelles, B-1050 Brussels,
Belgium; sgoriely@astro.ulb.ac.be}, Mathieu Samyn$^2$, 
and Naoki Itoh$^1$}

\bigskip
%\affil{The Astrophysical Journal, draft version (do not distribute!)}
%\affil{The Astrophysical Journal, Submitted 2003 November 11}
\affil{The Astrophysical Journal, Accepted 2004 January 20}

\begin{abstract}
The astrophysical origin of $r$-process nuclei remains a long-standing
mystery. Although some astrophysical scenarios show some promise, many
uncertainties involved in both the astrophysical conditions and in the
nuclear properties far from the $\beta$-stability have inhibited us from
understanding the nature of the $r$-process. The purpose of the present
paper is to examine the effects of the newly-derived microscopic
Hartree-Fock-Bogoliubov (HFB) mass formulas on the $r$-process
nucleosynthesis and analyse to what extent a solar-like $r$-abundance
distribution can be obtained. The $r$-process calculations with the
HFB-2 mass formula are performed, adopting the parametrized model of the
prompt explosion from a collapsing O-Ne-Mg core for the physical
conditions and compared with the results obtained with the HFB-7 and
droplet-type mass formulas. Due to its weak shell effect at the neutron
magic numbers in the neutron-rich region, the microscopic mass formulas
(HFB-2 and HFB-7) give rise to a spread of the abundance distribution in
the vicinity of the $r$-process peaks ($A = 130$ and 195). While this
effect resolves the large underproduction at $A \approx 115$ and 140
obtained with droplet-type mass formulas, large deviations compared to
the solar pattern are found near the third $r$-process peak. It is shown
that a solar-like $r$-process pattern can be obtained if the dynamical
timescales of the outgoing mass trajectories are increased by a factor
of about $2-3$, or if the $\beta$-decay rates are systematically
increased by the same factor.
\end{abstract}

\keywords{nuclear reactions, nucleosynthesis, abundances --- stars:
          abundances --- supernovae: general}

\section{INTRODUCTION}

The origin of the rapid neutron-capture ($r$-process) nuclei is still a
mystery. One of the underlying difficulties is that the astrophysical
site (and consequently the astrophysical conditions) in which the r-process 
takes place has not been identified. Although some scenarios such as ``neutrino-driven
winds'' from nascent neutron stars \citep{Woos94, Taka94, Qian96, Card97,
Otsu00, Wana01, Thom01}, ``prompt supernova explosions'' from small iron
cores \citep{Sumi01} or from O-Ne-Mg cores \citep{Wana03}, and ``neutron
star mergers'' \citep{Frei99b} show some promise, each of them faces
severe problems and cannot at the present time be called for to explain
the production and galactic enrichment of the $r$-process nuclei
observed in nature.

Another underlying difficulty is due to the uncertainties in the
theoretical predictions of nuclear data far from the $\beta$-stability,
for which essentially no experimental data exist. In particular, mass
predictions for neutron-rich nuclei play a key role since they affect
all the nuclear quantities of relevance in the $r$-process, namely the
neutron capture, photodisintegration and $\beta$-decay rates, as well as
the fission probabilities . Although most of the recent mass formulas
show fits to experimental masses of similar quality (characterized by an
rms error of about $0.7$~MeV), the mass extrapolations far from the
valley of $\beta$-stability can differ from each other quite
significantly \citep[for a recent review, see][]{Lunn03}.  Recently,
Hartree-Fock mass formulas with fully microscopic approaches have been
constructed \citep{Gori01a, Samy02,Gori02,Samy03,Gori03}. The latest
Hartree-Fock-Bogoliubov formula, labeled HFB-2 up to
HFB-7\citep{Gori02,Samy03,Gori03}, are among the most accurate mass
formulas, predicting the 2135 measured masses with a root-mean-square
error around 0.670~MeV for nuclei with $N,\,Z\ge 8$, i.e with the same
accuracy or even better than the one obtained with droplet-like mass
formulas \citep{Hilf76,Moel95}. A brief comparison of these mass models
is given in \S~2.

The purpose of this study is to examine the effects of the newly-derived
microscopic mass formula on the $r$-process nucleosynthesis and analyse
to what extent a solar-like distribution can be obtained. Most of
previous works devoted to the study of nuclear mass formulas and their
impact on the $r$-process nucleosynthesis were based on {\it
site-independent} approaches, the so-called canonical model assuming a
constant temperature and neutron number density during the
neutron-capture phase \citep{Gori96, Cowa99, Gori99, Gori01b,
Scha02}. This approach is understandable since the astrophysical
$r$-process site has not been unambiguously identified. Another approach
consists in considering promising sites and in modifying some of the
relevant characteristics to force a successful $r$-process. In this
case, the free parameter space is usually reduced and an easier analysis
can be performed. In addition, some works based on such {\it
site-specific} approaches have demonstrated that the abundance
distributions are sensitive to the astrophysical conditions adopted, in
particular during the ``freezeout'' phase \citep{Surm97, Frei99a,
Surm01, Wana02}.

We adopt, here, for the physical conditions the semi-realistic
astrophysical model of the ``prompt supernova explosion'' from the
collapsing O-Ne-Mg core by \citet{Wana03} (\S~3). The $r$-process
nucleosynthesis with the HFB-2 mass formula in each outgoing mass
trajectory is then calculated with a nuclear reaction network code. The
mass-averaged yields over the mass shells relevant for the $r$-process
is compared with the $r$-process abundance patterns in the solar system
and in an extremely metal-poor star (CS~22892-052), as well as with
those obtained with other mass formulas, more specifically the HFB-7
mass prediction and the extensively used droplet formulas of
\citet{Hilf76} and \citet{Moel95} (\S~4).  Uncertainties stemming from
the astrophysical conditions and extra nuclear ingredients are discussed
in \S~5. A summary follows in \S~6.

\section{MICROSCOPIC MASS MODELS}

Among the ground state properties, the atomic mass is obviously the most
fundamental quantity and influences the $r$-process abundance
predictions mainly through the $(n,\gamma)-(\gamma,n)$ competition
taking place in the neutron-rich region.  The
calculation of the reaction rates also requires the knowledge of other
ground state properties, such as the deformation, density distribution,
single-particle level scheme, pairing force and shell correction energies.
Recently, impressive progress has been made experimentally. This situation
results largely from recent measurements with Penning-trap or Schottky
spectrometers which have enlarged the region of known masses, in particular
towards the neutron-deficient side of the valley of nuclear stability
\citep[for a recent review, see][]{Lunn03}. The new Atomic Mass Evaluation
\citep{Audi01} contains 2214 measured masses, i.e 250 more than the 1995 one
\citep{Audi95}. A more accurate mass determination is also available now for about 132
nuclides originally included in the 1995 compilation. Out of the 382 new
experimental masses, 337 are located in the proton-rich region of the
nuclear chart and only 45 in the neutron rich region. As far as nuclei
directly involved in the $r$-process are concerned, almost no
experimental mass data exist, and theory must fill the gap.

Attempts to estimate nuclear masses go back to the liquid-drop
Weizs\"acker mass formula. Improvements to this original model have been
brought little by little, leading to the development of
macroscopic-microscopic mass formulas, such as the droplet model (DM)
\citep[e.g][]{Hilf76} and the `finite-range droplet model' (FRDM) of
\citet{Moel95}. In this framework, the macroscopic contribution to the
masses and the microscopic corrections of phenomenological nature are
treated independently, both parts being connected solely through a
parameter fit to experimental masses.  Despite the great empirical
success of these formulas (e.g. FRDM fits the 2135 $Z\ge 8$ experimental
masses \citep{Audi01} with an rms error of 0.676 MeV), it suffers from
major shortcomings, such as the incoherency of the link between the
macroscopic part and the microscopic correction, the instability of the
mass prediction to different parameter sets, or the instability of the
shell corrections. As a consequence, its reliability when extrapolating
far from experimentally known masses is severely limited. Although these
models have been extensively used for astrophysics applications
\citep{Woos94, Taka94, Sumi01, Wana01, Wana02, Wana03}, there is an
obvious need to develop a mass model that is more closely connected to
the basic nuclear interaction properties.

A new major progress has been achieved recently within the Hartree-Fock
method \citep{Gori01a,Samy02,Samy03,Gori02,Gori03}. It is now
demonstrated that this microscopic approach, making use of a Skyrme
force fitted to essentially all the mass data, is not only feasible, but
can successfully compete with the most accurate droplet-like formulas
available nowadays in the reproduction of measured masses. This holds
true not only when the pairing force is described in the BCS
approximation, but also when the Bogoliubov method is adopted (HFB
model), which has the advantage of ensuring the self-consistency of the
treatment of the nuclear single-particle and pairing properties. These
large-scale HFB calculations are based on the conventional 10-parameter
Skyrme force along with a 4-parameter $\delta$-function pairing
force. The Skyrme and pairing parameters are derived from a fit to the
full data set of 2135 measured masses of nuclei with $Z,N \geq 8$,
leading to an rms error of the order of 0.674~MeV for the HFB-2 mass
table.

Despite the success of the HFB-2 mass formula, a series of studies of
possible modifications to the basic force model and to the method of
calculation was initiated all within the HFB framework in order to test
the reliability of the mass extrapolations \citep{Samy03,Gori03}. For
this reason, a set of additional 5 new mass tables, referred to as HFB-3
to HFB-7 were designed to analyse the sensitivity of the mass fit and
extrapolations to the prescription used for the center-of-mass
correction, the density-dependence of the pairing interaction and the
amplitude of the effective nucleon mass.  The new mass fits are of the
same quality as the HFB-2 mass predictions.  In addition, it is found
that globally the extrapolations out to the neutron-drip line of all
these different HFB mass formulas are essentially equivalent. Figure~1
(lower panel) compares the HFB-2 and HFB-7 masses for all nuclei with
$8\le Z \le 110$ lying between the proton and neutron
driplines. Although HFB-2 and HFB-7 are obtained with significantly
different Skyrme forces (in particular, HFB-2 is characterized by an
density-independent pairing force and an effective isoscalar mass
$M^*_s=1.04$, while HFB-7 has an density-dependent pairing force and
$M^*_s=0.8$), deviations smaller than about 2 MeV are obtained for
nuclei with $Z \le 82$.

In contrast, more discrepancies are seen between HFB and droplet-type
masses (Figure 1), especially for superheavy nuclei. For lighter
species, the mass differences remain below some 5-10~MeV, although
significantly different shell and deformation effects are
predicted. Most particularly, the HFB mass formulas show a weaker
neutron-shell closure close to the neutron drip line with respect to
droplet-like models as FRDM \citep[for a more detailed discussion,
see][]{Gori02}. Future improved microscopic predictions, in particular
in mean field, but also shell model approaches, as well as future
experiments will hopefully shed light on the behavior of the $N = 82$
and 126 neutron shell closures at large isospins, in a similar way as
for the $N = 20$, 28, and 50 magic numbers \citep{Lunn03}. These major
differences in the nuclear structure properties may affect the
nucleosynthesis predictions, as shown in \S~4.

\section{SUPERNOVA MODEL AND THE $r$-PROCESS}

Recent comprehensive spectroscopic studies of extremely metal-poor stars
\citep{Hill02, Cowa02, Sned03}, as well as chemical evolution studies
\citep{Ishi99, Ishi04, Arga03}, in the Galactic halo suggest that the
$r$-process is a primary nucleosynthesis process and that the
astrophysics site could be associated with core-collapse supernovae. In
fact, most of the recent studies on the nucleosynthesis $r$-process have
been based on the ``neutrino-driven wind'' scenario, in which the free
nucleons accelerated by the intense neutrino flux near the neutrino
sphere of a core-collapse supernova assemble to heavier nuclei
\citep{Woos94, Qian96, Card97, Otsu00, Wana01, Thom01}.

In the present study, we use the parametrized model of the ``prompt
supernova explosion'' from an $8-10 M_\odot$ progenitor star (with a
$1.38 M_\odot$ O-Ne-Mg core) by \citet{Wana03}. The reason is that this
model leads to $r$-abundance distributions that have been shown to be
relatively similar to the solar distribution, at least if an artificial
enhancement of the shock-heating energy is assumed. In addition, this
scenario does not suffer from the problematic overproduction of $A
\approx 90$ nuclei seen in the neutrino-driven wind model \citep{Woos94,
Wana01}. Finally, the physical conditions in hydrodynamical (i.e.,
prompt) explosions may not be significantly affected by highly uncertain
neutrino heating (or cooling) processes \citep{Hill84} as they are in
the neutrino wind model. The site modeling is consequently easier and
the final abundance predictions less sensitive to the neutrino
physics. All these reasons justify the present study and the comparison
of the nucleosynthesis results with the solar $r$-abundance
distribution.  It is obvious, however, that a similar study based on the
neutrino-driven wind scenario is needed since the prompt explosion
scenario also faces severe problems that need to be resolved
\citep{Wana03}. In particular, it should be recalled here that in
consistent realistic prompt explosion models, only a weak explosion is
obtained and no $r$-processing. An artificial enhancement of the shock
energy is needed to drive an energetic explosion and provide the
necessary conditions for a successful $r$-process.

\citet{Wana03} performed core-collapse simulations with a one-dimension,
Newtonian hydrodynamic code. The highly neutronized ejecta ($Y_{e0}$,
initial electron mole fraction, $\approx 0.14-0.20$) in the (artificial)
energetic explosion (model~Q6) is subject to a strong production of
$r$-process nuclei. In this study, we use the mass trajectories of the
ejected material from $M_{\rm ej} = 0.08 M_\odot$ (zone number 60,
$Y_{e0} = 0.45$) to $0.31 M_\odot$ (zone number 105, $Y_{e0} = 0.14$) in
mass coordinate (the surface of the O-Ne-Mg core is at mass coordinate
zero), as shown in Figure~2. This mass range was shown by \citet{Wana03}
to give a solar-like $r$-abundance distribution assuming that part of
the ejected material with low-$Y_e$ \citep[layers 106 to 132 of ]
[]{Wana03} was artificially assumed to be reaccreted onto the compact
object, and consequently not to contribute to the galactic
enrichment. The temperature and density histories of some mass elements
are presented in Figure~3.

Adopting model Q6 in \citet{Wana03} for the physical conditions, the
$r$-process abundances are obtained by solving an extensive nuclear
reaction network code. The network consists of $\sim 4400$ species, all
the way from single neutrons and protons up to the fermium ($Z = 100$)
isotopes (Figure~4). We include all relevant reactions, i.e. $(n,
\gamma)$, $(p,\gamma)$, $(\alpha, \gamma)$, $(p, n)$, $(\alpha, n)$,
$(\alpha, p)$, and their inverse. All reaction rates are calculated
within the statistical model of Hauser-Feshbach making use of
experimental masses \citep{Audi01} whenever available or the HFB-2 mass
predictions \citep{Gori02} otherwise. The photodisintegration rates are
deduced from the reverse $(n,\gamma)$ rates applying the reciprocity
theorem with the nuclear masses considered. The $\beta$-decay and
$\beta$-delayed neutron emission rates are taken from the gross theory
(GT2) of \citet{Tach90}, obtained with the ETFSI \citep[extended
Thomas-Fermi plus Strutinsky integral,][]{Abou95} $Q_{\beta}$
predictions. Other nuclear inputs are the same as in \citet{Wana03}.

A word of caution is required concerning some inconsistencies in the
nuclear input considered here, and most particularly concerning the use of
ETFSI masses to estimate $\beta$-decay rates within the GT2 model. As
mentioned in \S~2, nuclear masses influence the $r$-abundance distribution
mainly through the $(n,\gamma)-(\gamma,n)$ competition taking place in the
neutron-rich region. Thanks to the balance theorem, such a competition
remains relatively independent of the model adopted to estimate the
neutron capture and photodisintegration rates. In contrast, the sensitivity
of the absolute value of the reaction and $\beta$-decay rates on
nuclear masses can depend significantly on the model considered. For example,
within the statistical GT2 model, the $\beta$-decay rate scale as the fifth
power of the $Q_{\beta}$ value, while in the continuum QRPA model 
\citep[e.g ][]{Borz03b}, the rates depend on the possible
spin-isospin excitation within the
$Q_{\beta}$ window but are not directly affected by the nuclear mass as
such.  Ultimately, for each mass model, a consistent estimate of the reaction
and $\beta$-decay rates based on the corresponding mass model predictions
for masses, but also deformation, single-particle
properties, \dots should be performed. In order to simplify here the
discussion on the impact of masses, but also to avoid the complicate model
dependence in the estimate of the reaction and $\beta$-decay rates, we will
restrict ourselves to consider in all cases only one set of $\beta$-decay
rates, namely the GT2 rates with ETFSI masses.

The calculation in each mass trajectory is started at $T_9 = 9$ (where
$T_9 \equiv T/10^9$~K). The initial composition is taken to be that of
the nuclear statistical equilibrium with the matter density $\rho$ at
$T_9 = 9$, and consists mostly of free nucleons and alpha particles. The
calculation is terminated after 10 seconds from the start in each mass
trajectory. Snapshots of the $r$-process calculation in the
mass-trajectory 99 ($1.11 M_\odot$ in mass coordinate and $Y_{e0} =
0.16$) are shown in Figure~4. The abundances (mole fractions) are color
coded, while the isotopes included in the reaction network are shown by
dots with the stable and meta-stable isotopes by larger dots. Contours
of constant two-neutron-separation energies divided by two ($S_{2n}/2$)
from 1 to 8~MeV are superimposed. The abundance curve as a function of
the mass number is also shown in each upper left panel.

The upper panel of Figure~4 shows the time slice when the
neutron-to-seed ratio, $Y_n/Y_h$, decreases to $\sim 1$, where $Y_n$ is
the abundance of free neutrons and
\begin{equation}
Y_h \equiv \sum_{Z>2, A} Y(Z, A)
\end{equation}
that of nuclei heavier than helium. At this time, the $r$-process flow
closely follows an iso-$S_{2n}$ curve given by $S_{2n}/2 = S_a^0 \approx
1.6$~MeV (red line), where
\begin{equation}
S_a^0 ({\rm MeV})
\equiv \left(34.075 - \log N_n + \frac{3}{2} \log T_9 \right)
\frac{T_9}{5.04},
\end{equation}
$N_n$ is the number density of free neutrons in units of cm$^{-3}$. The
condition $Y_n/Y_h = 1$ is referred to as the {\it neutron exhaustion}.

For the adopted $(n,\gamma)$ rates, a quasi-equilibrium between neutron
capture and photodisintegration continues until $\tau_\beta/\tau_n$
decreases to $\sim 1$ as can be seen in the middle panel of Figure~4,
where
\begin{equation}
\tau_\beta \equiv
\left[\frac{1}{Y_h}\sum_{Z>2, A} \lambda_\beta (Z, A)\, Y(Z, A) \right]^{-1}
\end{equation}
and
\begin{equation}
\tau_n \equiv
\left[\frac{\rho Y_n}{Y_h} \sum_{Z>2, A} \lambda_n (Z, A)\, Y(Z, A) \right]^{-1}
\end{equation}
are the averaged $\beta$-decay and neutron-capture lifetimes, and
$\lambda_\beta(Z, A)$ and $\lambda_n(Z, A)$ the $\beta$-decay and
neutron-capture rates of the nucleus $(Z, A)$, respectively. At this
time, the abundance distribution is still near the $S_a^0 \approx
2.6$~MeV line. We refer hereafter the condition $\tau_\beta/\tau_n = 1$
as {\it freezeout}, and the epoch between neutron exhaustion and
freezeout as the {\it freezeout phase}. This freezeout phase is of
particular relevance to determine the final abundance pattern as
discussed in \S~4. After the freezeout, the global abundance pattern is
not significantly affected anymore, although it can still be somewhat
smoothed out by photodisintegration and $\beta$-delayed neutron
emission, as can be seen in the lower panel of Figure~4
($\tau_\beta/\tau_n \approx 0.2$).

\section{IMPACT OF MASS PREDICTIONS ON THE $r$-PROCESS}

The final $r$-process yields are mass-averaged over all mass
trajectories between 1.08 and $1.30 M_\odot$ in mass coordinate. In
Figure~5a, the mass-averaged abundances (line) are compared with the
solar $r$-process abundance pattern \citep[dots,][]{Gori99a} that is
scaled to match the height of the third $r$-process peak. For
comparison, identical calculations were performed by replacing our
standard HFB-2 masses by the HFB-7 \citep[][Figure~5b]{Gori03}, FRDM
\citep[][Figure~5c]{Moel95}, and DM \citep[][Figure~5d]{Hilf76}
predictions.

Roughly speaking, an overall agreement between calculated and solar
abundance curves can be seen in Figures~5a-d for nuclei with $A =
100-200$.  Discrepancies common to all cases are, however, observed, in
particular the deficiencies for nuclei in the vicinity of the first
$r$-process peak ($A = 80$), the shifts with respect to the solar curve
of the second and third $r$-process peaks ($A = 130$ and 195) to lower
mass numbers ($\sim 3$ units in mass number) and the second to third
peak height which is significantly smaller than observed in the solar
distribution. We do not discuss further on the deficiencies near $A =
80$, since theoretically the results are highly dependent on the adopted
mass cut of the ejecta \citep[see \S~3 and][for more details]{Wana03}
and observationally there has been an increasing number of evidences
that the light ($A < 130$) $r$-process nuclei have a different origin
from the heavy ($A > 130$) ones \citep[e.g.,][]{Hill02, Sned03}. For the
shifts of the peak locations, as well as the second to third peak ratio,
it is conceivable that this deficiency does not originate from the mass
model, but rather from inadequate astrophysical or nuclear description,
since it is observed for all mass formulas. This point will be further
discussed in \S~5.

A few significant differences in the abundance patterns can, however, be
observed near the second and third peaks when use is made of the
Hartree-Fock models (HFB-2 and HFB-7) on one side and the droplet models
on the other side. First, the underproduction of nuclei at $A \approx
115$ and 140 is more pronounced with the FRDM (Figure~5c) and DM
(Figure~5d) masses than with the HFB-2 (Figure~5a) or HFB-7 (Figure~5b)
masses. Second, the abundances near $A = 130$ in the HFB cases are
spread out in contrast to what is observed in the solar
$r$-abundances. Third, the abundance curves near the third peak with the
HFB masses are widened and the valley at $A = 183$ as observed in the
solar $r$-distribution is significantly shifted to lower mass numbers.

These differences reflect the model properties of iso-$S_{2n}/2$ curves
shown in Figure~6, along which the $r$-process proceeds (see also
Figure~4).  Major local differences between the HFB (Figures~6a, b) and
the droplet masses (Figures~6c, d) are found near the neutron magic
numbers $N = 82$ and 126. The Hartree-Fock masses show weaker
shell-closures, i.e. smoother iso-$S_{2n}/2$ curves, at $N = 82$ and
126.  This reduced shell effect \citep[which actually is not totally
quenched at the neutron dripline in contrast to other predictions,
e.g][]{Pear96} is responsible for spreading the second and third
abundance peaks. In contrast, the large droplet shell effect is clearly
seen at $N = 82$ and 126 by the steep $S_{2n}/2$ character even close to
the neutron dripline and gives rise to sharp abundance peaks.

Figure~7 illustrates the formation process of the second and third peaks
with the HFB-2 masses in more details. The left panels show the
snapshots of the abundance distributions for the trajectory 90 ($1.15
M_\odot$ in mass coordinate and $Y_{e0} = 0.23$), and the right panels
for the trajectory 99 ($1.11 M_\odot$ in mass coordinate and $Y_{e0} =
0.16$) at the neutron exhaustion (top panels), freezeout (middle
panels), and later time (bottom panels). The trajectory 90 (left panels
in Figure~7) gives a major contribution to the second-peak
formation. The abundances are widely distributed already at neutron
exhaustion (left-top panel), owing to the weaker shell gaps at $N = 82$,
in particular for low $S_{2n}/2$. This results in a split of the second
abundance peak as seen in Figure~5a. The shift of the abundance
distribution to the heavier mass number during the freezeout phase is
small, although the abundance curve is significantly smoothed by
photodisintegration (and slightly by $\beta$-delayed emission). The
reason is that the freezeout takes place at a low $S_a^0$-value
($\approx 2.9$~MeV, left-middle panel in Figure~7).

The formation process of the third peak in the trajectory 99 can be seen
in the right panels of Figure~7. The $N = 126$ shell effect in the
neutron-rich region is not reduced as the $N = 82$ one, though it still
remains weaker than in the droplet approach. The splitting of the peak
abundances is not observed, but the abundance curve is significantly
broaden by photodisintegrations during the freezeout phase. The
freezeout takes place when $S_a^0$ is still low (2.6~MeV), and for this
reason the peak position at $A = 190$ is not shifted during the
freezeout phase.

Some of the deficiencies observed in the $r$-abundance distribution
obtained with HFB masses are seen to be cured when considering droplet
masses (Figure~5). In particular, the DM masses characterized by
isospin-independent $N=82$ and $N=126$ shell effects give rise to an
abundance curve which is seen to be globally in better agreement with
the solar pattern than do the HFB curves. It would be tempting to judge
the quality of the nuclear mass predictions on the basis of such a
comparison. However, it must be recalled here that such $r$-process
calculations in prompt explosion scenarios (like in any other site
studied so far) remain strongly affected by other input uncertainties in
the astrophysical modeling as well as extra nuclear ingredients. More
specifically, the total mass finally ejected (i.e not reaccreted onto
the proto-neutron star) remains unknown and strongly dependent on the
astrophysical modeling of the explosion. The same holds for the
dynamical timescale of the outgoing material. Finally, major
difficulties still exist in predicting reliably the $\beta$-decay
half-lives of exotic neutron-rich nuclei. These uncertainties are
studied in \S~5, where it is shown that a modification of either the
physical conditions of the explosion or the $\beta$-decay rates can
influence drastically the agreement with the solar $r$-distribution.

Finally, it could also be argued that the outcome of the prompt
explosion nucleosynthesis in one specific star might not need to match
the solar $r$-process pattern. Recent spectroscopic studies show that
the elemental abundance patterns between the second and third peaks of
$r$-process-enhanced ultra-metal-poor stars in the Galactic halo are in
excellent agreement with the solar $r$-process abundances. In
particular, the agreement extends to the third peak elements Os, Ir, and
Pt for HD~115444, HD~126238 \citep{Sned98}, BD~+17$^\circ$3248
\citep{Cowa02}, and CS~22892-052 \citep[][Figure~8]{Sned03}, and
CS~31082-001 \citep{Hill02}.  These observations led these authors to
the conclusion that there is only one single $r$-process site that
contributes to the production of heavy ($A > 130$) $r$-process nuclei
and that the corresponding distribution is similar to the solar
$r$-process pattern \citep{Sned96}. Instead of comparing the predicted
abundances with the solar content, we can therefore use such
observations and estimate to what extent the disagreements seen in
Figure~5 and discussed above are also reflected in the comparison with
the observed elemental distribution.  The comparison is shown in
Figure~8 for the most-studied, extremely metal-poor star CS~22892-052
\citep{Sned03}. Even in this case, it is difficult to discriminate
between the predictions obtained with the four different mass formulas
considered here. This is quite normal since the observed distribution is
reduced to some 30 elements (in comparison with some 130 nuclei in the
solar material). And above all, the observed pattern could well be
explained by the nuclear properties invariance in the $56 \le Z \le 76$
region rather than from specific astrophysics conditions as discussed in
\citet{Gori97b}.  It is therefore more informative to compare the
$r$-abundance prediction with the solar pattern, keeping in mind that
there is no stringent observational constraint that the final
distribution should be strictly solar. Nevertheless, future abundance
determinations of as many extremely metal-poor stars as possible will be
of particular importance to shed light on nucleosynthesis yields from
only one or a few supernova events.

\section{UNCERTAINTIES IN THE $r$-PROCESS CALCULATIONS}

In this section, we examine possible astrophysical and nuclear
modifications leading to a better agreement of the HFB-2 abundance curve
with the solar $r$-process pattern. Regarding the prompt supernova
explosions considered here, we do not expect the entropy per baryon of
the shocked material to be very different from the one deduced from the
simulation ($\sim 10$ in units of the Boltzmann constant), since many
previous simulations resulted in similar values \citep{Hill84,
Sumi01}. However, the dynamical timescale of the outgoing matter
(without changing the entropy) can significantly differ from model to
model. For example, the dynamical timescales of the mass trajectories
responsible for $r$-process in \citet{Sumi01} are typically a few times
shorter than those applied in this study \citep{Wana03}, because of
their different treatments of electron capture. It is also conceivable
that other effects, such as convection, asymmetric explosion, or reverse
shock from the outer envelope, cause an acceleration or a deceleration
of the outgoing material.

In order to test the impact of a change in the dynamical timescales, we
simply modify the density and temperature profiles of each trajectory,
so that $\rho' (t) = \rho (t/f_t)$ and $T' (t) = T (t/f_t)$, i.e the
dynamical timescale is multiplied by a factor of $f_t$. Figures~9 and 10
show the abundance curves for $f_t = 0.2, 0.5$ (fast), 1 (standard), 2,
3, and 5 (slow) obtained with the mass-trajectories 90 and 99, which are
responsible for the formation of the second and third peaks,
respectively. The abundance curves near the second (Figure~9) and third
(Figure~10) peaks are seen to agree fairly well with to the solar
$r$-process pattern when the timescale is multiplied by a factor of
$2-3$ (Figures~9d-e and Figures~10d-e), and to disagree for the fast
trajectories (Figures~9a-b and Figures~10a-b).

This temporal effect is due to the significant abundance evolution
during the feezeout phase for the slowly outgoing material, as shown in
Figure~11 (same as Figure~7, but for $f_t = 3$). At the neutron
exhaustion, the abundances are dispersed near the second (left-top
panel) and third (right-top panel) peaks, respectively, owing to the
weak shell gaps at $N = 82$ and 126 for low $S_{2n}/2$ ($\approx
2-3$~MeV), as in the standard case (Figure~7). The freezeout takes place
(middle panels), however, at much higher $S_a^0$ values (4.2 and 3.7~MeV
for the trajectories 90 and 99, respectively) than those in the standard
case (2.9 and 2.6~MeV, Figure~7). The reason is that the temperatures at
the freezeout are higher ($T_9 = 1.4$ and 1.2 for trajectories 90 and
99, respectively) owing to the slow expansion with $f_t = 3$ than those
in the standard case ($T_9 = 1.0$ and 0.93). As a result, the abundance
peaks shift to $A = 130$ (trajectory 90) and 194 (trajectory 99), i.e
mostly at the same position as those in the solar
$r$-distribution. Concomitantly, the width of the abundance curve near
the second and third peaks is reduced significantly due to the larger
shell gaps at $S_a^0 \approx 4$~MeV than that at $S_a^0 \lesssim 3$~MeV,
and becomes similar to the one observed in the solar pattern (Figures~9e
and 10e).

{\it Assuming all the nuclear inputs are correct and the $r$-process
indeed originates from the collapse of O-Ne-Mg core}, this would imply
that the expansion of the outgoing material must be significantly slower
than obtained by \citet{Wana03}. A possible explanation could be found
in the deceleration of the ejecta by the reverse shock from the outer
envelope if the hydrogen and helium layers survive the mass loss till
the onset of the core collapse. This effect is absent from the
simulation of \citet{Wana03} with a bare O-Ne-Mg white dwarf. This
reverse shock effect would also reduce the total overproduction of
$r$-processed material per event in the prompt explosion scenario
\citep[see][]{Wana03}. It is interesting to note that such a significant
deceleration of the outgoing material takes place in the
neutrino-powered explosion due to the stalled shock wall once launched
by a core bounce \citep{Woos94}. More detailed numerical simulations of
the collapsing O-Ne-Mg core including the outer envelope are needed to
quantify this effect.

The final mass-averaged abundance curve corresponding to mass
trajectories slowed by a factor of three ($f_t = 3$) is shown with the
scaled solar $r$-process abundances in Figure~12a. We find a good
agreement between the calculated and solar $r$-process patterns, in
particular near the second and third peaks. The steep valley at $A
\approx 183$ observed in the solar $r$-process abundances is also well
reproduced. However, an underproduction at $A \approx 115$ and 140
appear.

As far as nuclear uncertainties are concerned, $\beta$-decays rates,
neutron-captures rates and fission probabilities also play a key role
\citep[neutrino reactions are not important in the prompt explosion,
see][]{Wana03}. A change in their uncertain prediction can affect the
nucleosynthesis predictions. In the present scenario, fission does not
play an important role however, since it affects only a small fraction
of the total $r$-processed matter \citep{Wana03}. The prediction of
neutron capture rates, especially for exotic neutron-rich nuclei,
remains difficult, in particular in the prediction of the direct capture
contribution and of the $\gamma$-ray strength and neutron-nucleus
optical model potential for exotic neutron-rich nuclei as required in
the statistical reaction model of Hauser-Feshbach \citep[for more
details, see][]{Gori97, Gori03a}. Much work remains to be done to
estimate the neutron capture rates reliably. For this reason, their
impact on the $r$-process nucleosynthesis is postponed to a future
study.

Recent microscopic calculations of the $\beta$-decay rate within the
energy density functional plus continuum QRPA formalism shows that the
first forbidden transitions might have been underestimated in the past
and can increase the $\beta$-decay rate by a factor of about 2 along the
neutron-rich $N=82$ isotone and of about 3-10 along the neutron-rich
$N=126$ isotone compared with the calculation based exclusively on the
allowed transitions \citep{Borz03}. Detailed microscopic estimates of
the $\beta$-decay rates (including forbidden transitions) for all nuclei
of relevance in $r$-process calculations, as well as future experiments
with radioactive ion beams are deeply awaited to improve this
fundamental input in the description of the $r$-process nucleosynthesis.

To estimate the influence of $\beta$-decays, we show in Figure~12b the
mass-averaged $r$-process yields obtained by multiplying all the
$\beta$-decay rates by a factor of three (i.e., reducing $\tau_\beta$ by
a factor of three). Interestingly, no significant differences are seen
between Figures~12a and 12b. For $\beta$-decay rates faster by a factor
of three, the freezeout (corresponding to $\tau_\beta = \tau_n$) takes
place at higher temperatures and thus at higher $S_a^0$ value, which has
globally the same effect as slowing down the outgoing material by the
same factor. This test calculation shows the crucial role played by
$\beta$-decay rates when considering non-parametrized density and
temperature profiles, i.e thermodynamic conditions fixed by
(semi-)realistic models.

\section{SUMMARY}

We have examined the $r$-process nucleosynthesis with the
state-of-the-art microscopic mass formula, HFB-2 \citep{Gori02}, to see
its effects on the theoretical $r$-process abundance prediction. The
yields of $r$-process species were calculated with a nuclear reaction
network code ($\sim 4400$ isotopes), adopting favorable physical
conditions in the prompt explosion model of a collapsing O-Ne-Mg core by
\citet{Wana03}. This model led to production of the solar-like
$r$-abundance pattern in the previous study \citep{Wana03}. However, an
artificial enhancement of the shock-heating energy was needed to obtain
requisite physical conditions for successful $r$-processing.

Due to its weak shell effect at the neutron magic numbers in the
neutron-rich region, the microscopic mass formulas (HFB-2 and HFB-7)
give rise to a spread of the abundance distribution in the vicinity of
the $r$-process peaks ($A = 130$ and 195). While this effect resolves
the large underproduction at $A \approx 115$ and 140 obtained with
droplet-type mass formulas, large deviations compared to the solar
pattern are found near the third $r$-process peak. When use is made of
droplet mass predictions, sharp $r$-process peaks are systematically
found, owing to their strong shell effect for neutron magic numbers even
in the neutron-rich region. However, due to the numerous uncertainties
still affecting the astrophysics models as well as the prediction of
extra nuclear ingredients, it would be highly premature to judge the
quality of the mass formula on the basis of such a comparison.

We found that abundance peaks similar to the one observed in the solar
system could be recovered if the dynamical timescales of the mass
trajectories are increased by a factor of $\sim 2-3$ (without any change
in the entropy) or by decreasing systematically the $\beta$-decay
half-lives by the same factor. These changes might be conceivable when
considering the current uncertainties in the astrophysics as well as in
the nuclear $\beta$-decay model. Much effort in the astrophysics and
nuclear modeling remain to be devoted to improve the difficult
description of the $r$-process nucleosynthesis.

\acknowledgments

This work was supported by a Grant-in-Aid for Scientific Research
(13640245, 13740129) from the Ministry of Education, Culture, Sports,
Science, and Technology of Japan. M.S. and S.G. are FNRS Research Fellow
and Associate, respectively.

\clearpage

\begin{figure}
%\epsscale{0.7} 
%\plotone{f1.eps} 
\includegraphics[height=15cm]{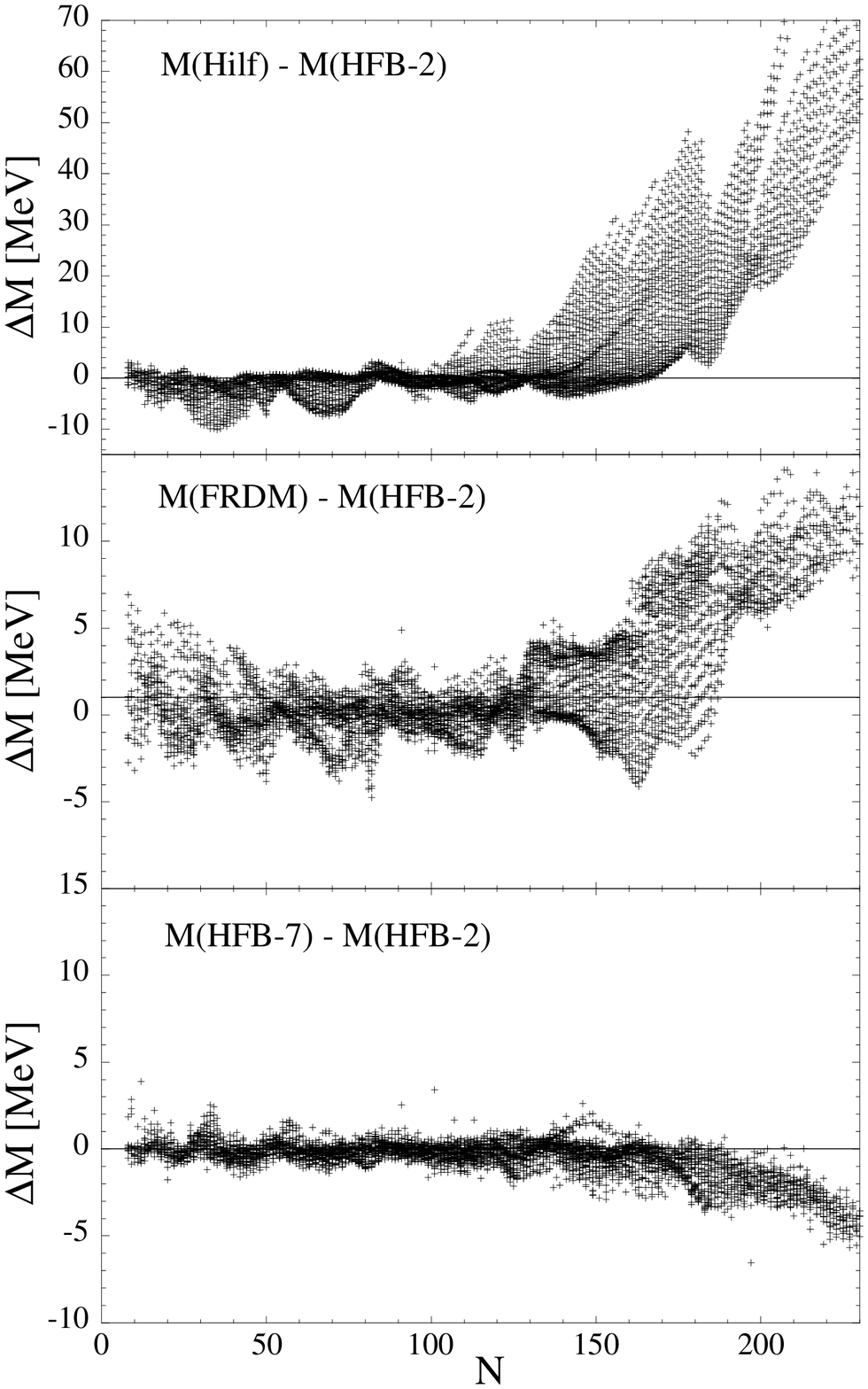}
\caption{Mass differences between HFB-2 and the DM of \citet{Hilf76}
(upper panel), HFB-2 and the FRDM of \citet{Moel95} (middle panel) and HFB-2
and HFB-7 (lower panel) as a function of the neutron number
$N$ for all nuclei with $Z, N \ge 8$ lying between the proton and the
neutron driplines up to $Z=110$}
\end{figure}

\clearpage

\begin{figure}
%\epsscale{1.0}
%\plotone{f2.eps}
\includegraphics[width=0.5\textwidth]{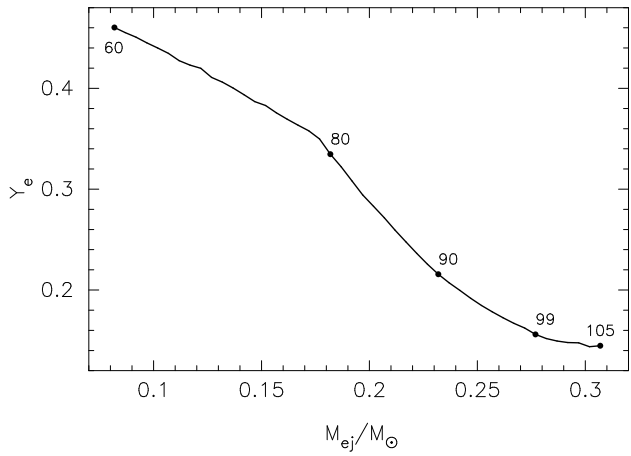}
\caption{Distribution of the initial electron mole fraction in the
ejected matter adopted in this study. The surface of the O-Ne-Mg core is
at mass coordinate zero. Selected mass elements are denoted by zone numbers.}
\end{figure}

%\clearpage

\begin{figure}
%\epsscale{1.0}
%\plotone{f3.eps}
\includegraphics[width=0.5\textwidth]{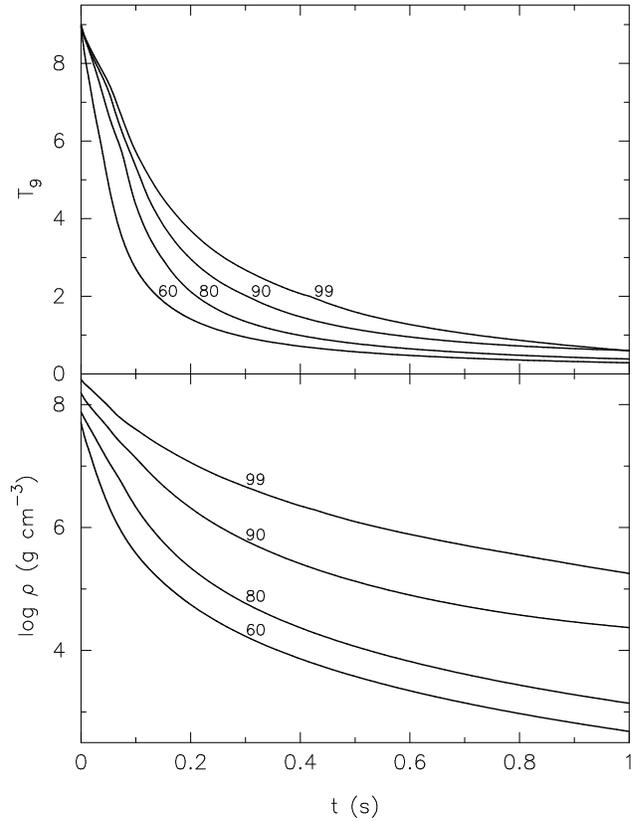}
\caption{Temperature (upper panel, $T_9
\equiv 10^9$~K) and density (lower panel) histories for selected mass
elements, which are denoted by zone numbers. Time is set to zero at $T_9
= 9$.}
\end{figure}

\clearpage

\begin{figure}
%\epsscale{0.6} 
%\plotone{f4.eps} 
\includegraphics[width=10cm]{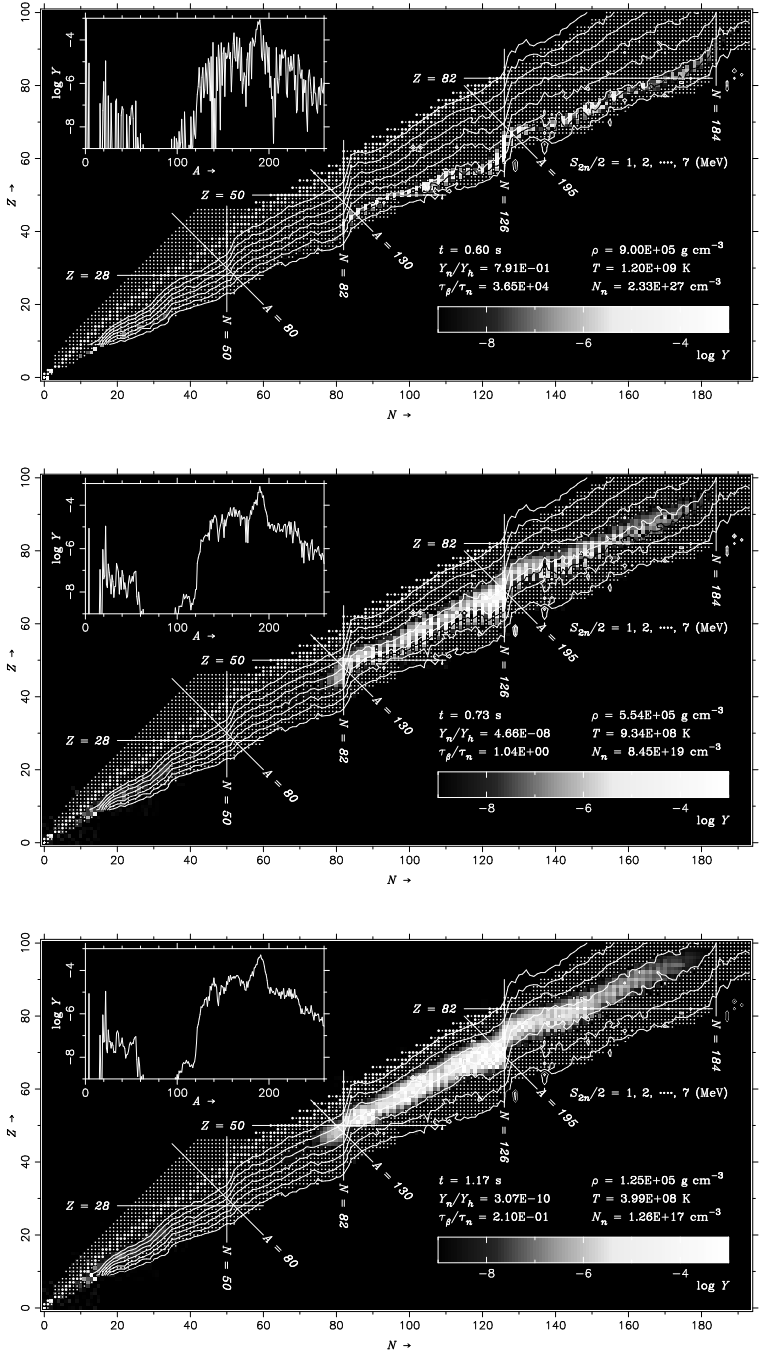}
\caption{Abundance distributions at the neutron exhaustion
($Y_n/Y_h \sim 1$, top), freezeout ($\tau_\beta/\tau_n \sim 1$, 
middle), and later time (bottom) for the
trajectory 99 (see text). The abundances are color-coded in the nuclide chart.
The abundance curve as a function of mass number is shown
in the upper left of each
panel. The nuclei included in the reaction network are denoted by dots, with
the stable and long-lived isotopes represented by large
dots. Iso-$S_{2n}/2$ (= 1, 2, $\cdots$, 7~MeV) and $S_a^0$
curves are also superimposed.}
\end{figure}

\clearpage

\begin{figure}
%\epsscale{1.0} 
%\plotone{f5.eps} 
\includegraphics[width=\textwidth]{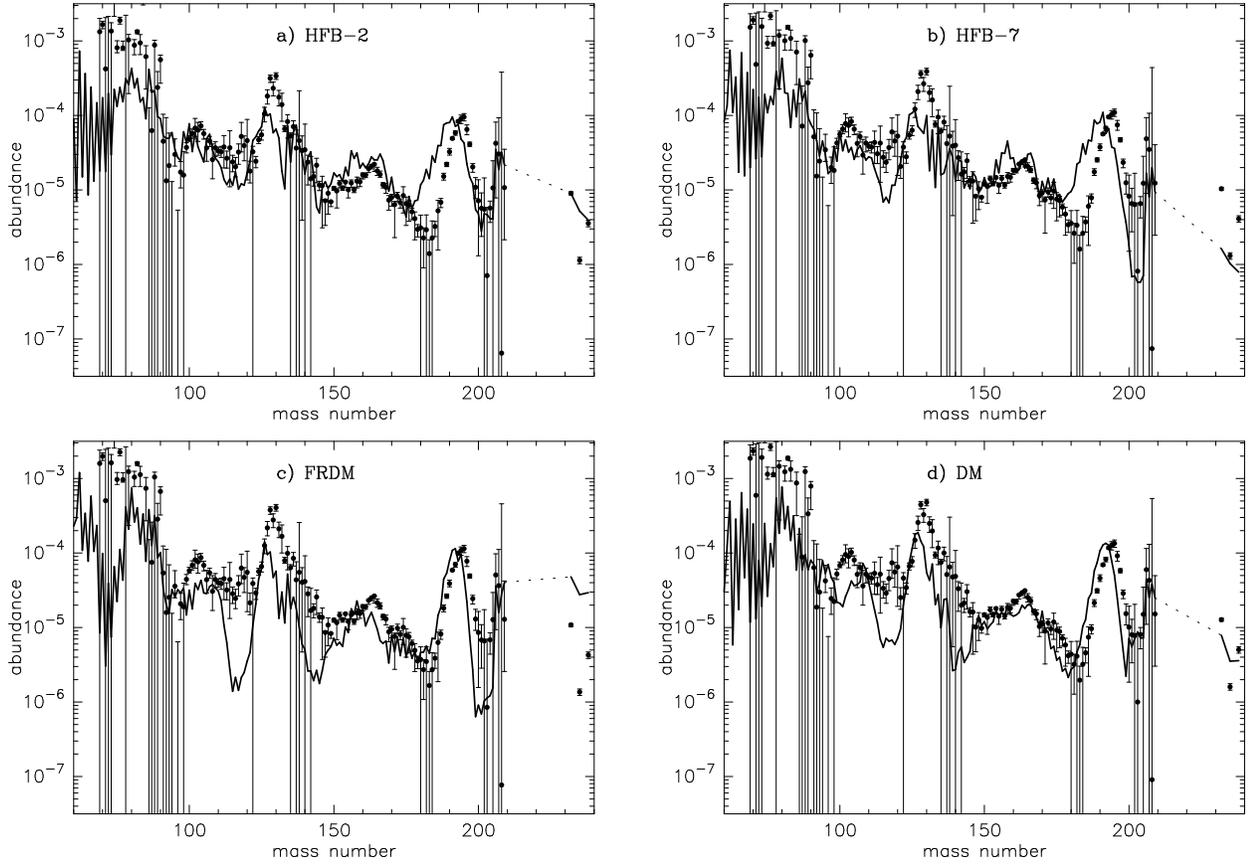}
\caption{ Final mass-averaged
$r$-process abundances (line) as a function of mass number obtained with
various mass formulas; (a) HFB-2, (b) HFB-7, (c) FRDM, and (d) DM.
These are compared with the solar $r$-process
abundances (points) from \citet{Gori99a}, which are scaled to match the
height of the third $r$-process peak.}
\end{figure}

\clearpage

\begin{figure}
%\epsscale{1.0} 
%\plotone{f6.eps} 
\includegraphics[width=\textwidth]{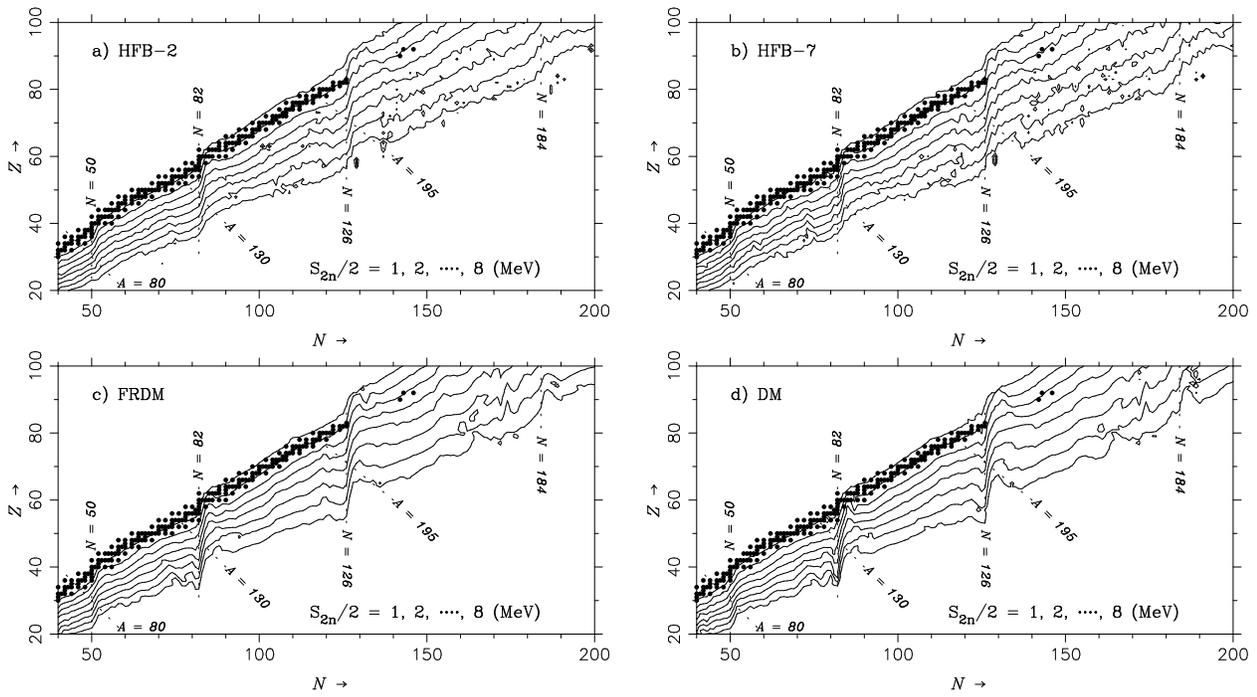}
\caption{Contours of the $S_{2n}/2$ values (= 1, 2, $\cdots$, 8~MeV) for
various mass formulas; (a) HFB-2, (b) HFB-7, (c) FRDM, and (d) DM.
The stable and long-lived isotopes are also represented by
dots.
}
\end{figure}

\clearpage

\begin{figure}
%\epsscale{0.9} 
%\plotone{f7.eps} 
\includegraphics[width=0.9\textwidth]{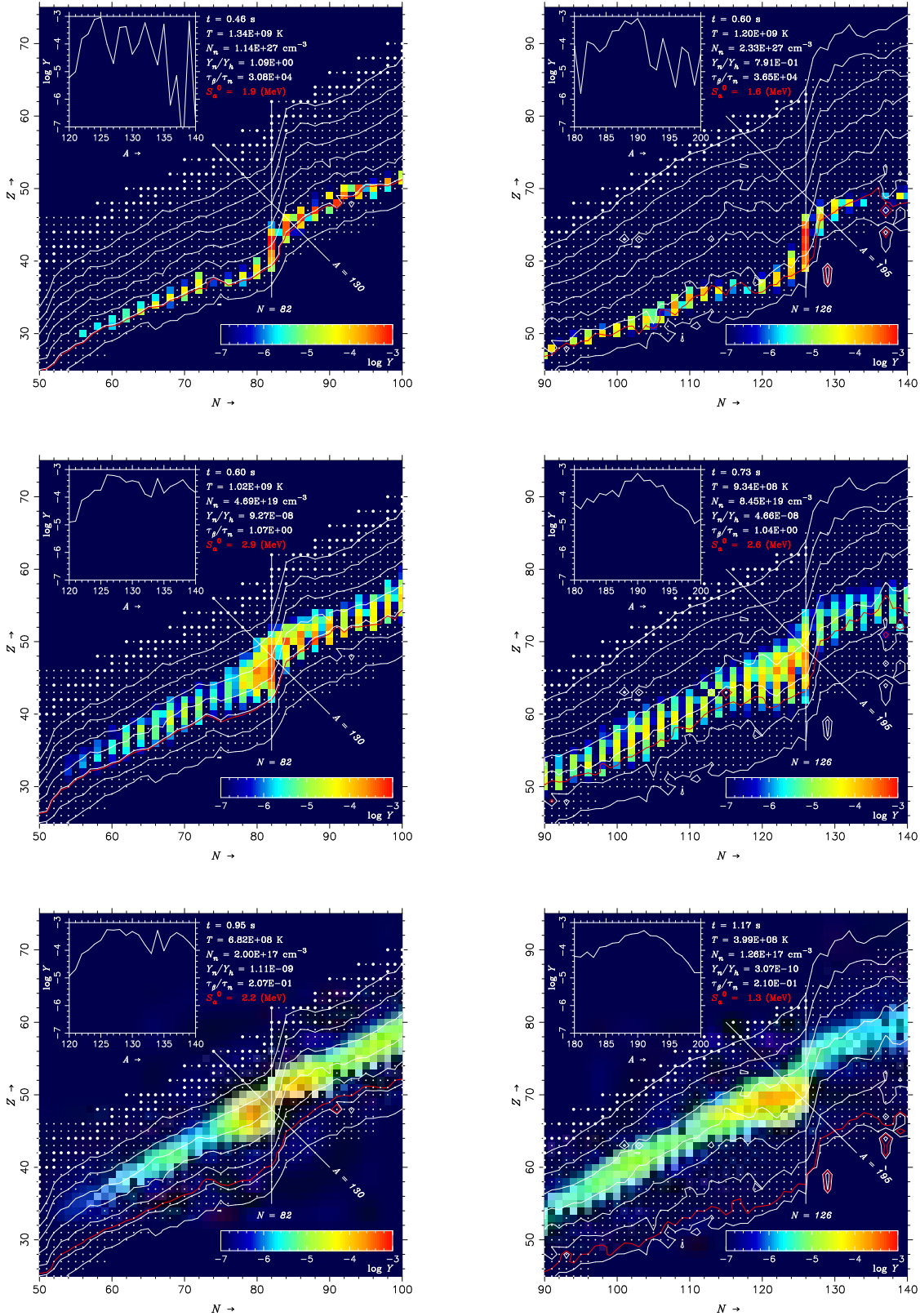}
\caption{Same as Figure~4, but near the second peak
($A \approx 130$) with the trajectory 90 (left
panels) and the third peak ($A \approx 195$)
with the trajectory 99 (right panels). }
\end{figure}

\clearpage

\begin{figure}
%\epsscale{1.0} 
%\plotone{f8.eps} 
\includegraphics[width=\textwidth]{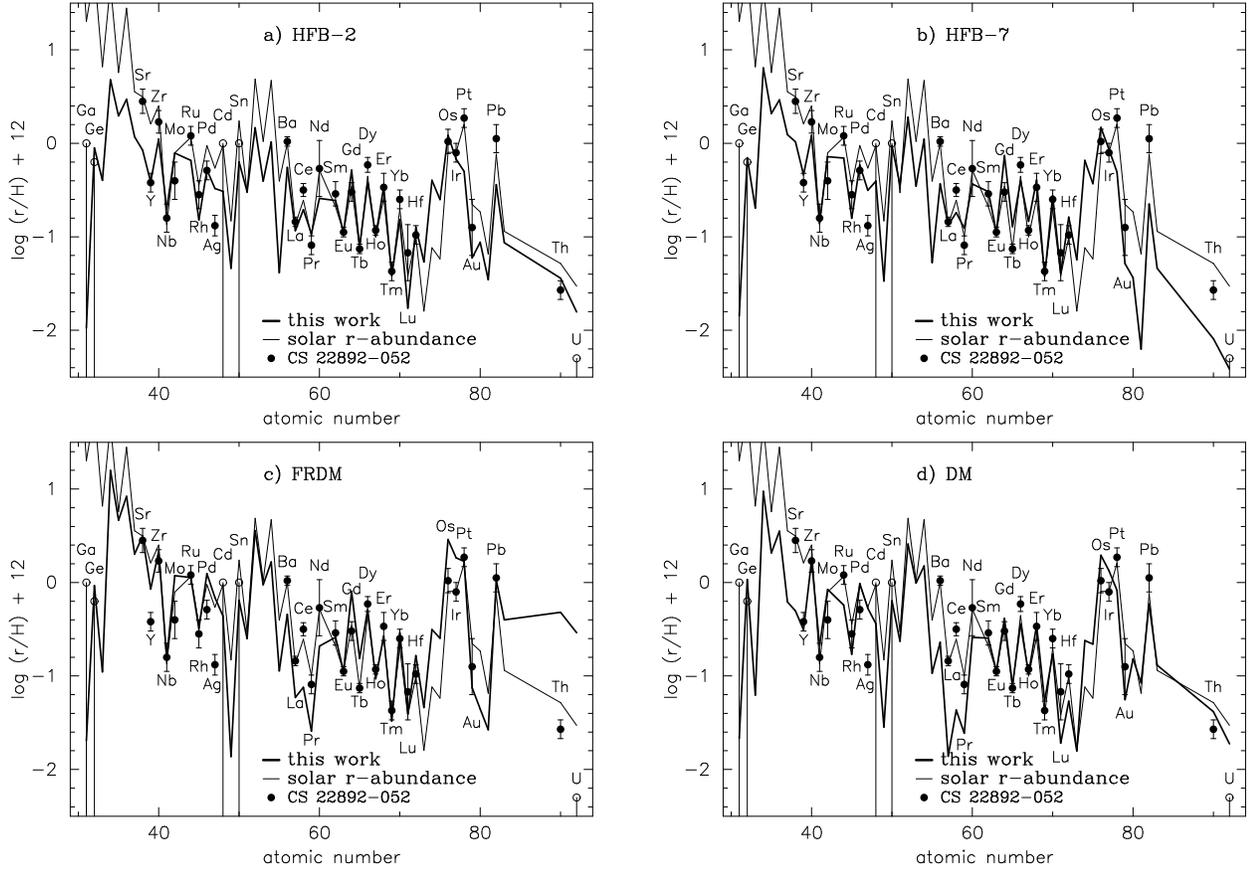}
\caption{Comparison of the
mass-averaged yields (thick line) obtained with the HFB-2 masses, scaled at Eu
($Z = 63$), with the abundance pattern of CS~22892-052 (filled circles, with
observational error bars), as a function of atomic number. For Ga, Ge,
Cd, Sn, and U, the observed upper limits are shown by the open circles. The
scaled solar $r$-process pattern is shown by the thin line.}
\end{figure}

\clearpage

\begin{figure}
%\epsscale{1.0} 
%\plotone{f9.eps} 
\includegraphics[width=\textwidth]{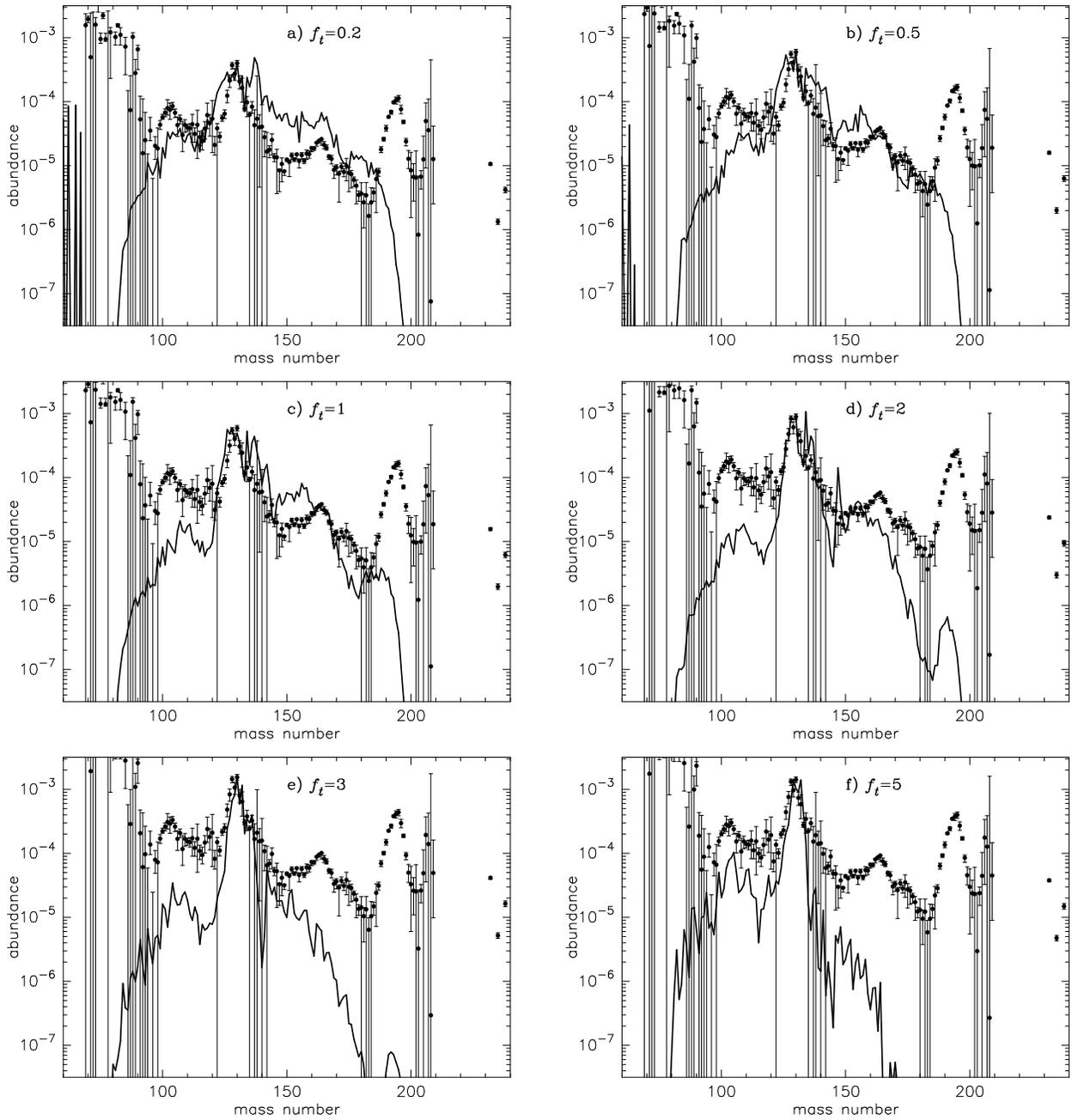}
\caption{Final abundances as a function
of mass number from $r$-process calculations for the trajectory 90
with $f_t =$ (a) 0.2,
(b) 0.5, (c) 1, (d) 2, (e) 3, and (f) 5 (see text).
These are compared with the solar $r$-process
abundances (points) from \citet{Gori99a}, which are scaled to match the
height of the second $r$-process peak.}
\end{figure}

\clearpage

\begin{figure}
%\epsscale{1.0} 
%\plotone{f10.eps} 
\includegraphics[width=\textwidth]{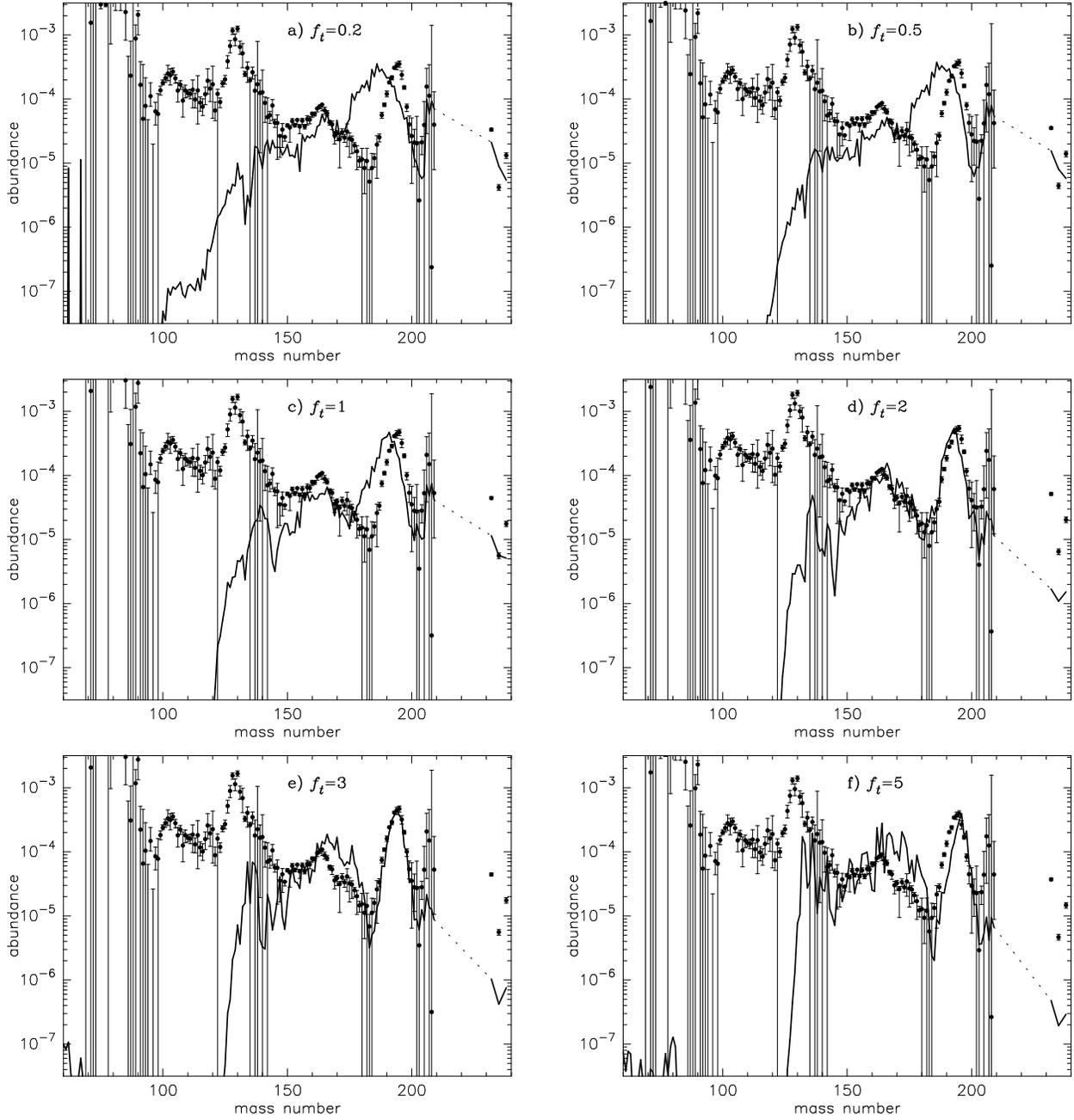}
\caption{Same as Figure~9, but for the trajectory 99. The solar $r$-process
abundances (points) from \citet{Gori99a} are scaled to match the
height of the third $r$-process peak.}
\end{figure}

\clearpage

\begin{figure}
%\epsscale{0.9} 
%\plotone{f11.eps} 
\includegraphics[width=0.9\textwidth]{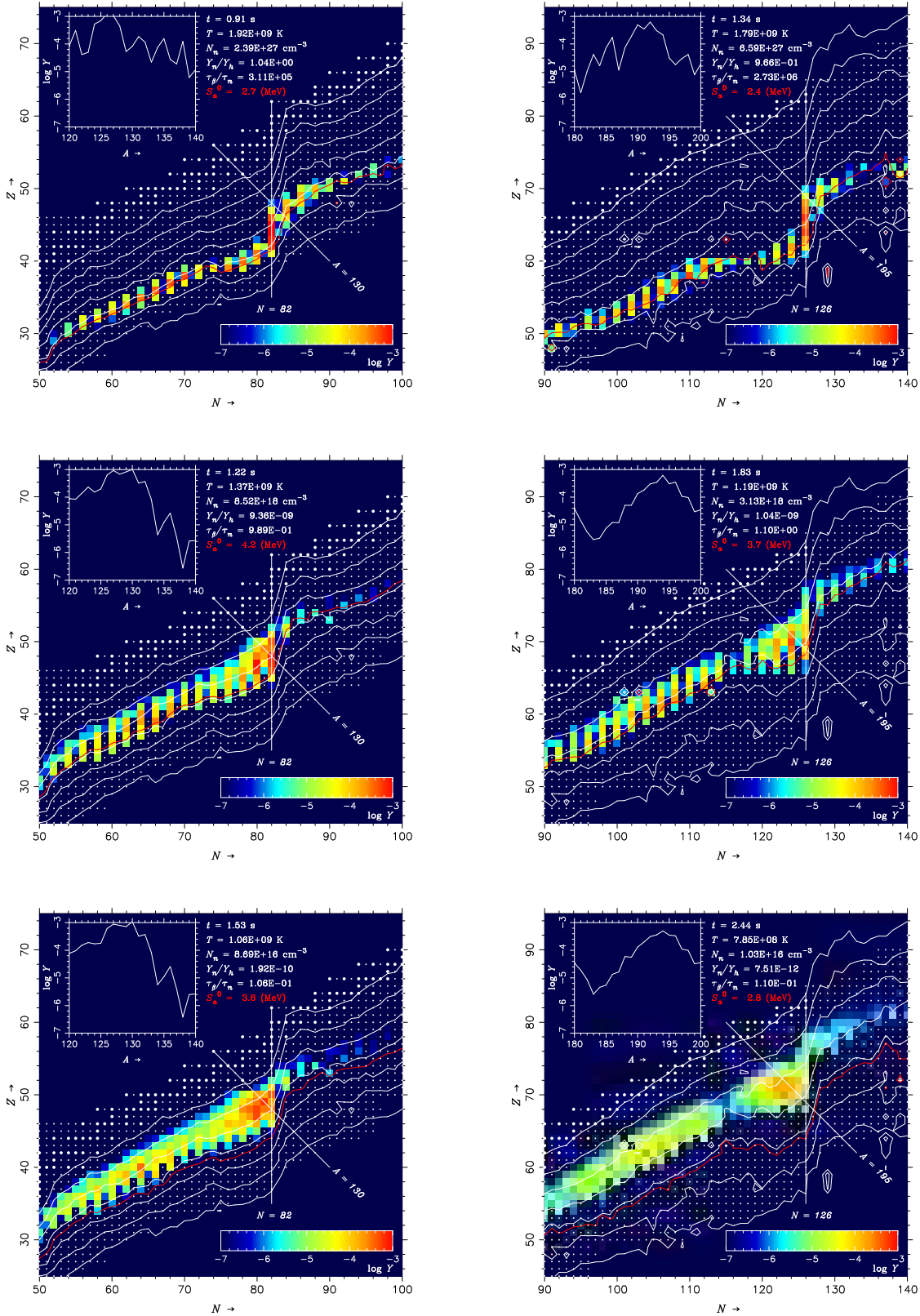}
\caption{Same as Figure~7, but for $f_t = 3$ (see text).}
\end{figure}

\clearpage

\begin{figure}
%\epsscale{1.0} 
%\plotone{f12.eps} 
\includegraphics[width=\textwidth]{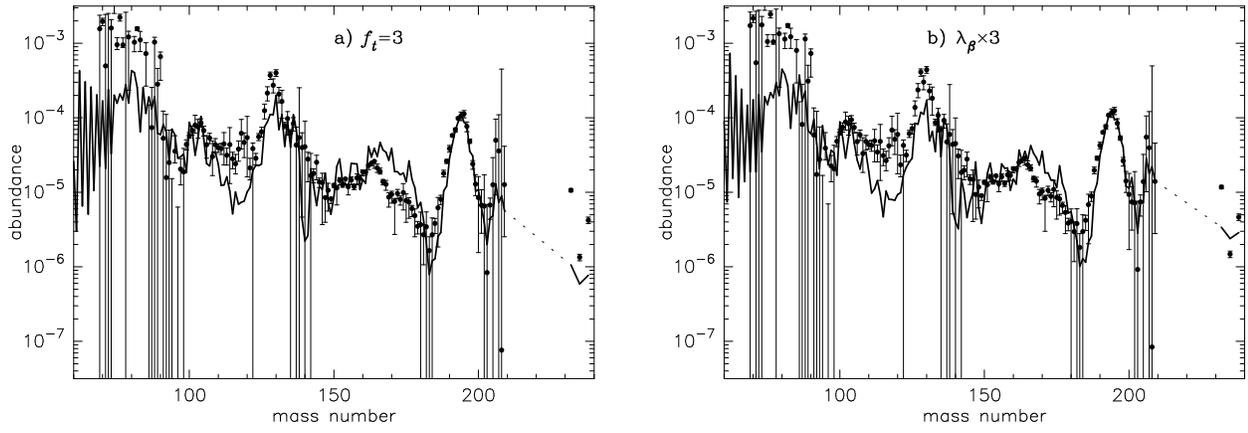}
\caption{Same as Figure~5, but for (a) slow trajectories ($f_t = 3$)
and (b) fast $\beta$-decay rates (a factor of three, see text).}
\end{figure}

\end{document}